\documentclass{emulateapj}

\newcommand{\til}{$\sim$}
\newcommand{\ergsqcmsec}{\thinspace\hbox{$\hbox{erg}\thinspace\hbox{cm}^{-2}
                \thinspace\hbox{s}^{-1}$}}
\newcommand{\ergsqcmsecA}{\thinspace\hbox{$\hbox{erg}\thinspace\hbox{cm}^{-2}
                \thinspace\hbox{s}^{-1}\thinspace\hbox{\AA}^{-1}$}}

\def\spose#1{\hbox to 0pt{#1\hss}}
\def\simlt{\mathrel{\spose{\lower 3pt\hbox{$\mathchar"218$}}
     \raise 2.0pt\hbox{$\mathchar"13C$}}}
\def\simgt{\mathrel{\spose{\lower 3pt\hbox{$\mathchar"218$}}
     \raise 2.0pt\hbox{$\mathchar"13E$}}}

\newcommand{\ta}{SDSS~J0729}
\newcommand{\tb}{SDSS~J0752}
\newcommand{\tc}{SDSS~J1700}
\newcommand{\sat}{{\em XMM-Newton}}

\newcommand{\II}{{\scriptsize~II}}

\newcommand{\VI}{{\scriptsize~VI}}
\newcommand{\VII}{{\scriptsize~VII}}

\def\today{November 3, 2004}

\slugcomment{Accepted for publication in the Astrophysical Journal \today}

\shorttitle{Polars from the SDSS}
\shortauthors{Homer et al.}

\begin{document}


\title{{\em XMM-Newton} and optical follow-up observations of three new Polars from the Sloan Digital Sky Survey$^{1,2}$}


\altaffiltext{1}{Some of the results presented here were obtained with the MMT
Observatory, a facility operated jointly by The University of Arizona and the
Smithsonian Institution.}
\altaffiltext{2}{Based on
observations obtained with the Sloan Digital Sky Survey and with the
 Apache Point
Observatory (APO) 3.5m telescope, which are owned and operated by the
Astrophysical Research Consortium (ARC)}
\author{Lee Homer\altaffilmark{3}, Paula Szkody\altaffilmark{3}, Bing Chen\altaffilmark{4}, Arne Henden\altaffilmark{5,6}, Gary D.
  Schmidt\altaffilmark{7},  Oliver J. Fraser\altaffilmark{3}, Karla Saloma\altaffilmark{3}, Nicole M. Silvestri\altaffilmark{3}, Hilda
  Taylor\altaffilmark{3}, and J. Brinkmann\altaffilmark{8}} 
\altaffiltext{3}{Department of Astronomy, University of Washington, Box 351580, Seattle, WA 98195, USA}
\altaffiltext{4}{\sat\ Science Operations Centre, ESA/Vilspa, 28080, Madrid, Spain}
\altaffiltext{5}{US Naval Observatory, Flagstaff Station, P.O. Box 1149, Flagstaff, AZ 86002-1149, USA}
\altaffiltext{6}{Universities Space Research Association}
\altaffiltext{7}{The University of Arizona, Steward Observatory, Tucson, AZ 85721, USA}
\altaffiltext{8}{Apache Point Observatory, P.O. Box 59, Sunspot, NM 88349-0059}
\email{homer@astro.washington.edu}




\begin{abstract}
We report follow-up \sat\ and optical observations of three new polars found in the Sloan Digital Sky Survey. Simple modeling of the X-ray spectra, and
consideration of the details of the X-ray and optical lightcurves corroborate the polar
nature of these three systems and provide further insights into their accretion characteristics.  During the \sat\ observation of
SDSS~J072910.68+365838.3, X-rays are undetected apart from a probable flare event, during which we find both the typical hard X-ray
bremsstrahlung component
 and a very strong O\VII\ ($E=0.57$~keV) line, but no evidence of a soft blackbody contribution.  In SDSS~J075240.45+362823.2 we identify an X-ray eclipse at
the beginning of the observation, roughly in phase with the primary minimum of the optical broad band curve.  The X-ray spectra require the presence of both hard and soft X-ray components, with their luminosity ratio
consistent with that found in other recent \sat\ results on polars.  Lastly,  SDSS~J170053.30+400357.6 appears optically as a very typical
polar, however its large amplitude optical modulation is $180^\circ$ out of phase with the variation in our short X-ray lightcurve.
\end{abstract}

\section{Introduction}
\begin{deluxetable*}{lllcll}[!tb]
\tableheadfrac{0.08}
\tablewidth{0pt}
\tabletypesize\small
\tablecaption{Observation Summary\label{tab:obslog}
}
\tablehead{\colhead{SDSS J}&
\colhead{UT Date} & \colhead{Obs} & \colhead{UT Time} & \colhead{Approx. V} &
\colhead{Comments} }
\startdata
0729&2002 Jan 07 & NOFS & 05:06 -- 10:49 &19.8--20.7 & open filter photometry\\
&2002 Oct 31 & APO: DIS &  10:40 -- 11:10   & 20.6 &spectrum \\
&2002 Oct 31& \sat: EPIC-pn & 20:39 -- 22:16 &\nodata &5282s live time\tablenotemark{a} \\
 && \sat: EPIC-MOS1/2 & 20:16 -- 22:21 & \nodata &7358/7392s live time \\
 && \sat: OM & 20:25 -- 22:26 & 20.7 &5999s duration, $B$ filter\\
&2003 Sep 22 & SO: SPOL & \nodata& \nodata&1800s, spectropolarimetry, pol.$=-2.79\%$\\
&2003 Nov 01 & MMT: SPOL &  \nodata& \nodata&1600s, spectropolarimetry, pol.$=+0.76\%$\\&2004 Feb 16 & SO: SPOL &  \nodata&\nodata& 2880s, spectropolarimetry, pol.$=+0.34\%$\\

0752    &2002 Oct 31 & NOFS & 08:33 -- 13:03 & 17.8--19.1&open filter photometry\\
&2002 Oct 31 & APO: DIS &  11:22 -- 11:37 & 17.5 &spectrum \\
&2002 Oct 31 & \sat: EPIC-pn & 23:40 -- 01:17 & \nodata &4019s live time \\
&\hspace*{7mm}--Nov 01 & \sat: EPIC-MOS1/2 &  23:18 -- 01:22 & \nodata &5990s live time \\
 && \sat: OM & 23:27 -- 01:28 & \nodata &6002s duration, $UVW2$ filter \\
&2003 Nov 01 & MMT: SPOL &  \nodata&\nodata& 2400s, spectropolarimetry, pol.$=-0.75\%$\\

1700&2003 Aug 11    & \sat: EPIC-pn & 16:57 -- 19:29 & \nodata &2659s live time \\
 && \sat: EPIC-MOS1/2 & 16:34 -- 17:07 &  \nodata &3546/3561s live time \\
 && \sat: OM & 17:53 -- 19:34 & \nodata &data lost\\
&2003 Jul 02 & NOFS & 03:42 -- 10:48& 17.9--19.5&open filter photometry\\
&2003 Aug 08 & MRO & 05:09 -- 11:38& \nodata &open filter photometry\\
&2003 Aug 09 & MRO & 04:44 -- 11:11& \nodata &open filter photometry\\
&2003 Aug 10 & MRO & 04:44 -- 10:57& 18.5--19.7&$B$ filter photometry\\

\enddata
\tablenotetext{a}{The live time of the X-ray CCD detectors refers to the sum of the good-time intervals, less any dead time.  It
  is typically much less than the difference of observation start and stop times.}
\end{deluxetable*}

The polars (or AM Her stars) possess the strongest magnetic fields ($B\sim10-200$MG) amongst Cataclysmic Variables (CVs).  In these systems, the plasma transferred
from the secondary does not form an accretion disc, but instead  the gas is threaded onto the field
lines and accretes at the poles on the white dwarf (WD) surface.  In the simplest picture, a strong shock develops in the accretion column, above
the surface, as the flow transitions from its high (approximately free-fall) velocity to a subsonic flow that can settle on the WD.  The
post-shock flow is a strong source of hard X-rays, mostly emitted as a thermal Bremsstrahlung continuum (with $kT_{\rm Br}=10-50$keV), although
line emission can also be important.  Further cooling occurs via cyclotron emission in the optical/IR, but is typically an order of magnitude
smaller.  Half of the hard X-ray photons  will be incident on the WD photosphere, heating it, so that it is a source of soft blackbody
emission (with  $kT_{\rm BB}=20-40$keV). For an X-ray albedo $a_X=0.3$ \citep{will87},  and neglecting the cyclotron contribution, the soft-hard energy
balance is then expected to be $L_{\rm BB}/L_{\rm
  Br}\sim0.56$ \citep{king87}.  However, observations of polars found in some cases (but not all) a large discrepancy with $L_{\rm BB}/L_{\rm
  Br}\simgt5$  \citep[see][and references therein]{rams94}, which was termed the ``soft X-ray excess.''  The explanation appears to lie in the
details of the accretion flow, it is both inhomogeneous, consisting of high and lower density regions, and likely blobby.  The former leads to
a hard X-ray deficiency, since the bulk of this emission arises from the highest density region, which may involve a small fraction of the total
flow.  The latter leads to enhanced blackbody.  In addition to the irradiation heating of the WD photosphere, if the flow is blobby the
longest/most dense blobs of material are able to penetrate deeply, before being shocked, and hence the energy released is thermalized before
emission. Lastly, at the very lowest specific accretion rates the flow degenerates into a bombardment solution \citep{kuij82}, where there are no
accretion shocks and 
the heating of the WD photosphere is also sufficiently weak that we expect no X-ray emission. For a review of the various accretion regimes see \citet{wick00}.

The Sloan Digital Sky Survey (SDSS) is in its fourth year of a multi-color photometric imaging and spectroscopic survey, which will eventually cover
\til25\% of the celestial sphere \citep{abaz03,abaz04,abaz04b,fuku96,gunn98,hogg01,lupt99,lupt01,pier03,smit02,stou02,york00}.  
Given an imaging limit of $g=23.3$ , SDSS is uncovering an unprecedented number of faint, low-accretion rate CVs amongst the $>150$ new systems discovered to date
\citep[see][]{szko02,szko03,szko04a}. Indeed, this is the very population that is expected to dominate CVs, but due to selection
biases was missed previously \citep{howe97}.

Several polars have been identified in the SDSS, among them two of the lowest accretion rate systems known \citep{szko03,szko04b} and the
shortest period eclipsing system \citep{schm04}.  \sat\ observations have helped to identify their accretion characteristics.
Here we report follow-up {\em XMM-Newton} X-ray and optical observations of three additional sources, SDSS J072910.68+365838.3,  075240.45+362823.2,
170053.30+400357.6 (hereafter SDSS J0729, 0752, 1700 for brevity), identified from their initial SDSS spectra as probable polars. Our
observations delineate the accretion regimes via measurement of their hard versus soft X-ray fluxes, and confirm the polar nature of the first two.  We note that
follow-up spectro-polarimetry of \tc\ reported in \citet{szko03} has already confirmed its polar nature. 

\begin{figure*}[!tb]
\resizebox{.95\textwidth}{!}{\rotatebox{0}{\plottwo{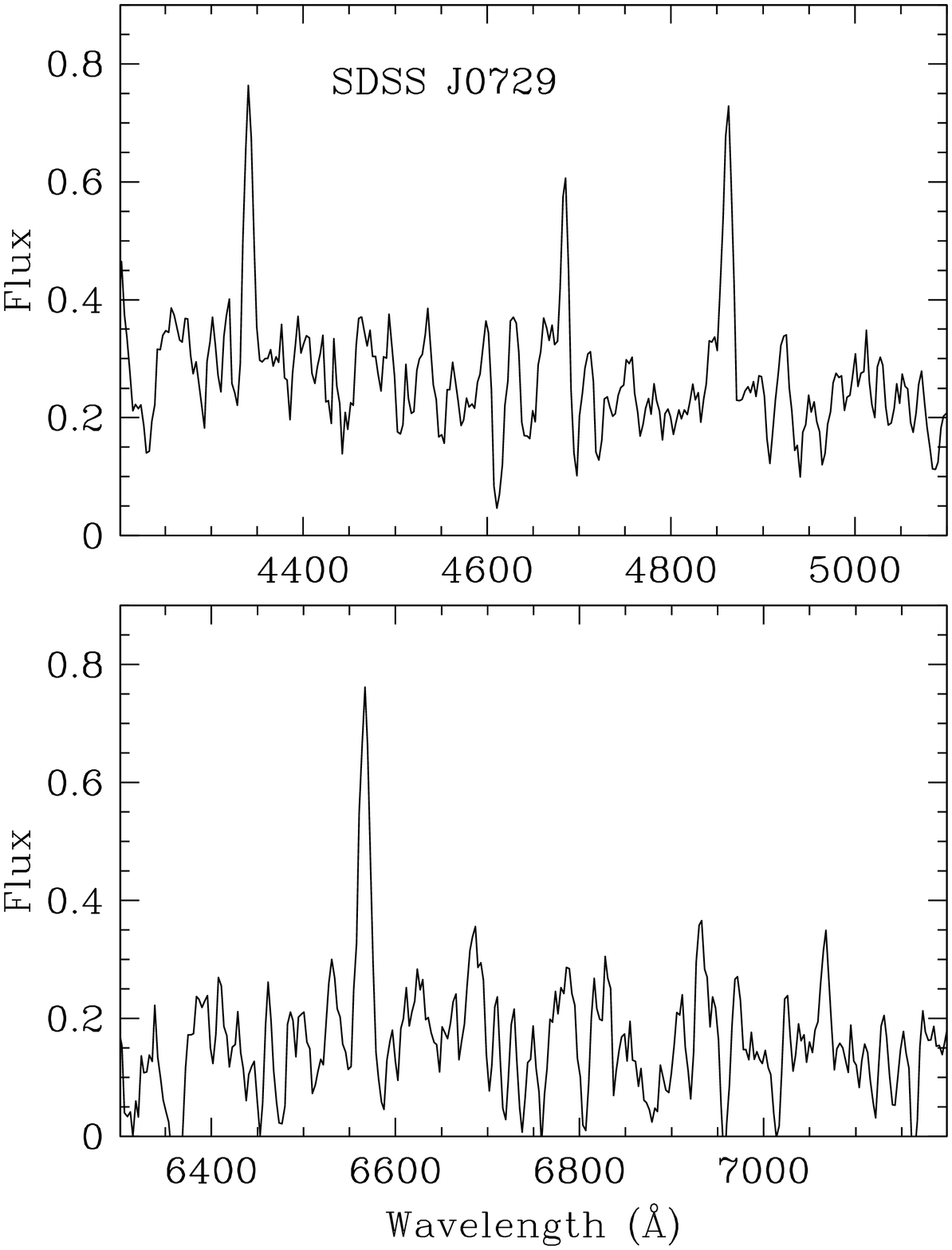}{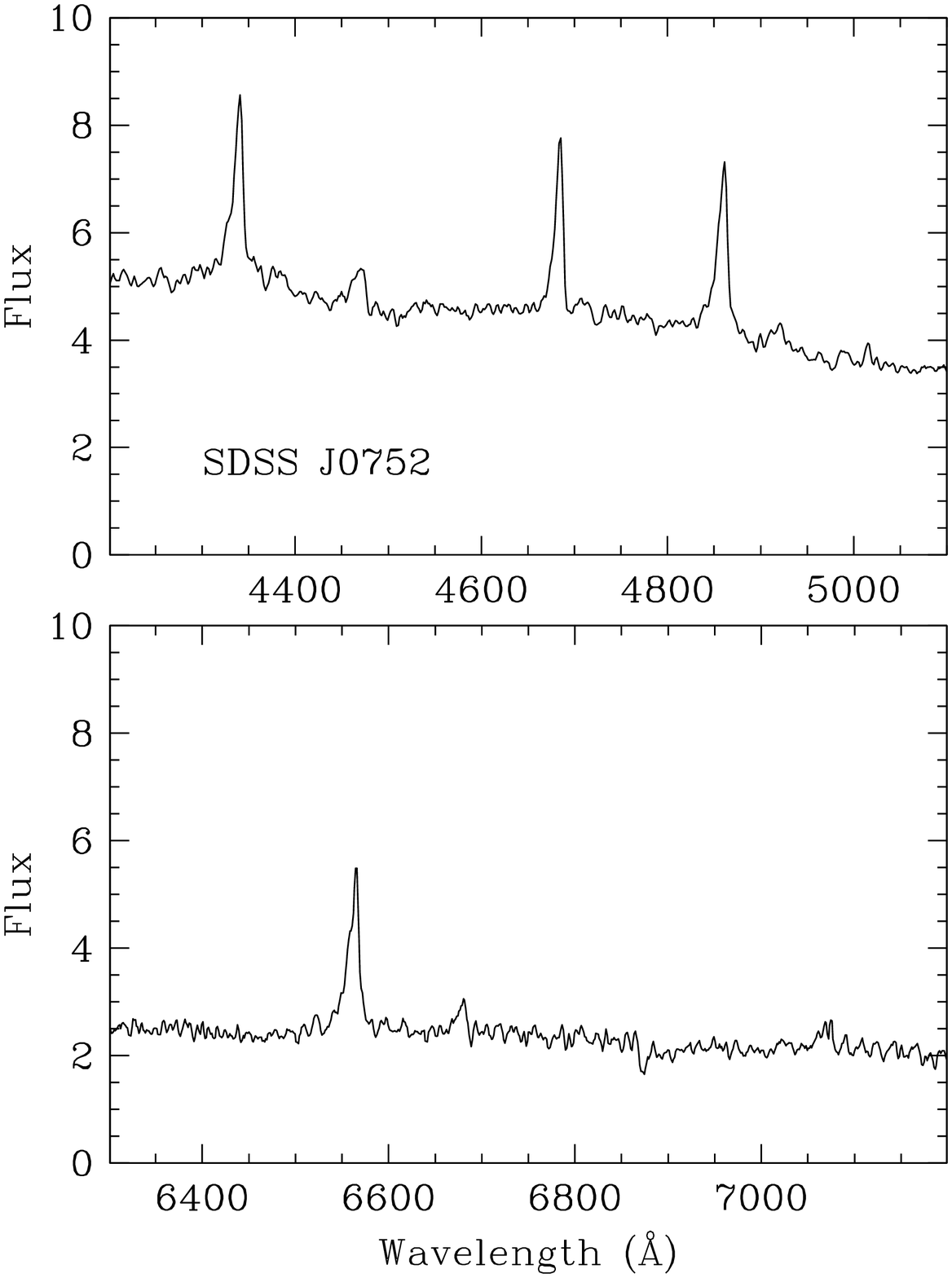}}}
\caption{Low-resolution spectra of \ta\ and \tb.  Both show strong He\II$\lambda4686$, which is characteristic of actively accreting
  polars. However, the emission lines of \ta\ are narrower than those of \tb\ and lack any indication of structure.  This indicates that the
  accretion rate of \ta\ was low at this time.  The flux scale is in units of flux density $10^{-16}$\ergsqcmsecA. \label{fig:optspec}}
\end{figure*}

\section{Observations}
For each \sat\ observation data are obtained with all detectors.  However, for our targets there were never sufficient counts to render the dispersed
spectra from the Reflection Grating Spectrograph \citep{denH01} of any use; even for the case of \tb, our brightest target, the continuum is barely
detected and no emission lines stand out. Optical Monitor \citep[OM,][]{maso01} data were successfully obtained
for only \ta\ and \tb.  However, in the UVW2 ultraviolet filter (required to avoid field brightness limits) \tb\ was not detected; data were obtained for
\ta\ using the $B$ filter.  Low-resolution spectra were available from the EPIC
camera; two MOS detectors \citep{turn01} + the pn \citep{stru01}, where the pn has roughly twice the effective area of each of the MOS. The UT times,
length of total observation and CCD livetime are listed in Table~\ref{tab:obslog}.  These data were reduced according to the guidelines from the
main \sat\ website (Vilspa\footnote{Available from
  http://xmm.vilspa.esa.es/external/xmm\_sw\_cal/ sas.shtml}) and also from the  US GOF ABC
guide\footnote{http://heasarc.gsfc.nasa.gov/docs/xmm/abc/abc.html}, using calibration files current to 2004 March 23 and the SAS v6.0.  Given
the calibration updates since pipeline processing, as a precaution we used SAS-tools to produce new event list files from the Observation Data
Files.  To check on variations in the non-X-ray background we created lightcurves for each entire detector in the 10--15~keV range, and where
appropriate created new GTIs to exclude intervals of higher background. We also screened these event lists using the standard canned expressions
and we restricted energies to the 0.1--10~keV range.   For the two MOS detectors, we used a 320 pixel radius circular aperture size (enclosing $\sim70\%$ of
the energy, chosen to maximize S/N), with a source-free background annulus surrounding the source;
  event pattern$\leq12$ selection was also applied.
For the pn, we extracted source data within a 360 pix (again encircled energy$\sim70\%$)
and background data from adjacent rectangular regions at similar detector Y locations to the target.  As advised, a conservative
  choice of event selection, pattern = 0, was adopted. 

\begin{figure}[!tb]
\resizebox{.49\textwidth}{!}{\rotatebox{0}{\plotone{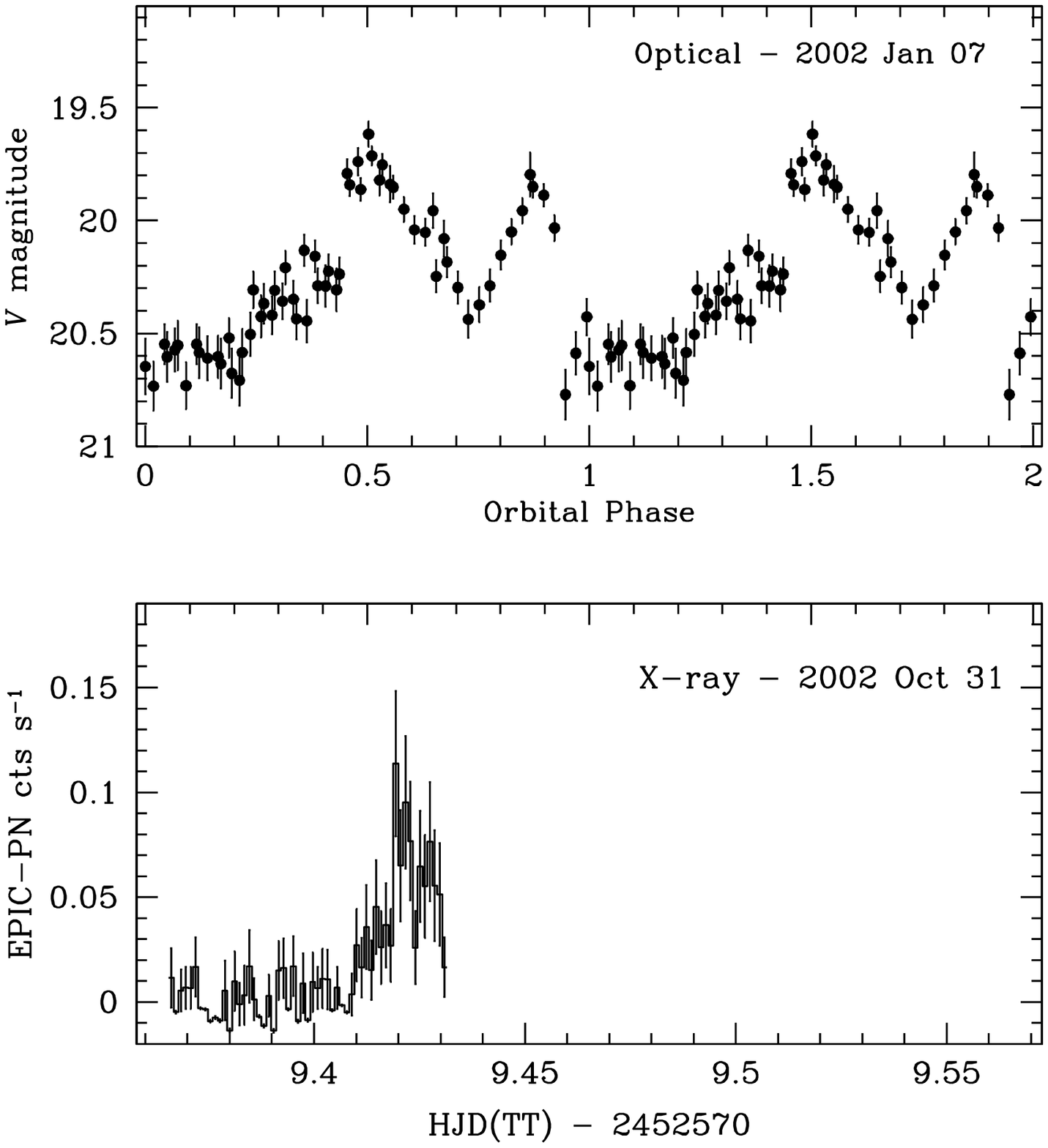}}}
\caption{Optical and X-ray lightcurves of \ta. {\em Top:} Orbital phase-folded $V$ band lightcurve. {\em Bottom:} 100s binned lightcurves from
  EPIC pn. An ephemeris was not available for the X-ray observations, but the scale is chosen to enable direct comparison to the optical curve
  above.  Note HJD(UT)=HJD(TT)--64.184 s at this epoch. \label{fig:0729lcs}}
\end{figure}

We fit simple blackbody (BB) + thermal Bremsstrahlung (Br) (or MEKAL\footnote{See
  http://heasarc.gsfc.nasa.gov/lheasoft/xanadu/xspec/ manual/node39.html} to include line emission) models to the various spectra to
obtain estimates of the relative contributions of soft ($\simlt0.6$~keV) and hard ($\simgt0.6$~keV) components.  Given the lack of counts from
\ta\ and the need to time-resolve the data from \tb, we chose to use Cash statistics \citep{cash79}, which work better for the case of few counts per
bin.  For \tc\ the background was a significant contribution (30--40\%), hence we opted to use background subtracted spectra and $\chi^2$
statistics for fitting.  In no cases were we able to reliably constrain the absorbing columns.  Hence, by default we adopted the value for the
position on the sky given by the HEASARC tool {\tt nH}, based on dust maps.  All our targets are at high Galactic latitude, leading to values
$N_H\sim$ few$\times10^{20}$cm$^{-2}$ typical of the values measured in most polars.  However, we do caution that our adopted values could be overestimates since these targets  are likely within a
few hundred parsecs of the Sun, and {\tt nH} gives the column to the edge of the Galaxy.

Lastly, in our fitting we used two sets of energy ranges (specifically the lower energy cut-offs), since the reliability of the calibrations at
the lowest energies is still unclear.  The large range (LR) set included energies $>$0.15keV for pn and $>$0.2keV for MOS, potentially useful
for constraining any soft blackbody component.  The restricted range (RR) had more conservative cut-off limits of $E>0.3$keV (pn) and $>$0.5~keV (MOS).

We extracted light curves for both source and background  with SAS task {\tt evselect}, using the same extraction regions as the spectra, but a
less conservative pattern$\leq4$ for the pn events.  Using
FTOOLS\footnote{http://heasarc.gsfc.nasa.gov/lheasoft/ftools/} tasks we subtracted the scaled background and converted the time stamps from
JD(TT) to HJD(TT)\footnote{The tools actually yield barycentric Julian Date in the barycentric dynamical time system, BJD(TB).  However, the
  offset to heliocentric Julian Date in the geocentric (terrestrial) dynamical time system (HJD(TT)) is less than \til3~s at any given time, fine for our
  purposes here.}

Contemporaneous optical spectra for \ta\ and \tb\ were obtained on 2002 October 31 using the double-imaging spectrograph (DIS) on the 3.5m telescope at Apache Point
Observatory (APO).  These were reduced and flux-calibrated within IRAF\footnote{IRAF (Image Reduction and Analysis Facility) is distributed by the National Optical Astronomy Observatories, which are operated
  by the Association of Universities for Research in Astronomy (AURA) Inc., under cooperative agreement with the National Science Foundation}.
In addition, differential photometry was obtained for \ta\ (in 2002 January), and \tb\ (contemporaneously), with the
US Naval Observatory Flagstaff Station (NOFS) 1m telescope; see Table~\ref{tab:obslog} for details.   These data were taken with no filter to give maximum
signal-to-nose, but an approximate zeropoint for Johnson $V$  was made possible through calibration of the field from all-sky photometry
including Landolt standards observed at NOFS. Hence, the lightcurves are labeled $V$ magnitudes, but actually show broad, white light results. Lastly, optical spectrophotometry and
circular spectropolarimetry were obtained for \ta\ and \tb\ in 2003 September, November and 2004 February with either the 2.3~m Steward Observatory
(SO) Bok
telescope or the 6.5~m MMT atop Mt.
Hopkins (see Table~\ref{tab:obslog}).  Both runs made use of the CCD SPOL spectropolarimeter \citep{schm92}. Neither of the sources was found to be highly polarized (magnitude of polarization$<1.0\%$ is considered a non-detection), only the 2003 September 22 observation of \ta\ made a
detection with a polarization of $-2.79\%$.  However, we note that \tb\ was in a low state during its only observation on 2003 November 01.

SDSS~J1700 was also observed by the NOFS 1m on 2003 July 2, and then, close to the \sat\ observations, by the 0.76m telescope at Manastash Ridge
Observatory (MRO) on 2003 August 8--10.  Time-resolutions \til$200$s
were achieved on all nights using no filter, apart from the final MRO night which used a $B$ filter. Details of the observations are listed in Table~\ref{tab:obslog}.  Differential magnitudes were
determined relative to at least three local standard stars in the field.

\section{Results}
From our contemporaneous optical spectra and/or lightcurves we are able to confirm that each of the targets were in their normal active accretion
states at the times of the \sat\ X-ray observations.  The low-resolution spectra of \ta\ and \tb\ show the characteristic He II$\lambda4686$
(Fig.~\ref{fig:optspec}), and are in general very similar to
their SDSS spectra.  The NOFS photometry of \tb\ also gave a consistent $V=18-19$.  Likewise \tc\ was observed at
$B=19.33\pm0.03$, a similar flux level to all prior optical observations.

\subsection{SDSS J0729}
In the optical, repeatable variability is apparent, with two full cycles covered.  From a PDM analysis we derive a period of $2.5\pm0.1$
hr.  Though of very low S/N, the $V$ light curve from the simultaneous OM observations shows consistent morphology.   In Figure~\ref{fig:0729lcs} we
show the folded lightcurve and also the \sat\ lightcurve for an equivalent time interval.  The complex optical morphology is reminiscent of that seen in AM Her on occasion \citep[see e.g.][]{maze86}.  \citet{gans01} have recently
developed a quantitative model for the $B$ and $V$ lightcurves of AM Her. The accretion column, responsible for the optical cyclotron emission, is
always visible, with the primary minimum occurring when the line-of sight is along the accretion funnel, and the secondary minimum due to
partial self-eclipse by the WD and/or cyclotron beaming effects. 
\begin{figure}[!tb]
\resizebox{.48\textwidth}{!}{\rotatebox{0}{\plotone{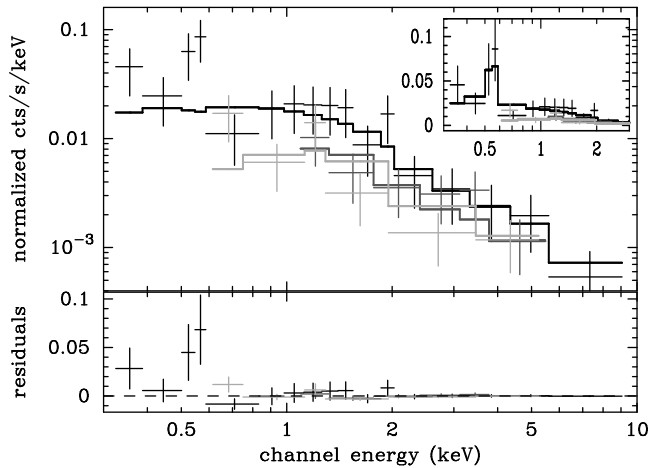}}}
\caption{EPIC spectra of \ta\ for the X-ray 'flare' event (pn: black lines, MOS1: light grey, MOS2: dark grey). The best fit bremsstrahlung model is over-plotted and residuals are shown below, the
  excess at \til0.57~keV is clear.  In the inset we show the bremsstrahlung + Gaussian line fit. \label{fig:0729xspec}}
\end{figure}

The X-ray flux in \ta\ is undetectable for fully 0.4 in
phase, then brightens significant, possibly for only \til0.2.  This is very similar to the X-ray flare behaviour seen in a low state of UZ For
\citep{pand02}, suggesting that the accretion rate is sufficiently low (consistent with the optical spectrum) that any X-ray emitting region is too cool to fall within the \sat\\ passband most of the
time.  However, given our incomplete phase coverage it is also possible that we are simply seeing the X-ray emitting part
of the column coming out of self-eclipse by the WD.  Since the X-ray emitting region is typically smaller and closer to the WD surface than the
cyclotron, a much longer X-ray self-eclipse is certainly plausible.

The spectrum from this high X-ray flux interval is also quite unusual.  We fit a Br + BB model to the restricted range; we constrained the temperatures to 30~keV and 40~eV respectively, as typical values for polars \citep{warn95}.  The result indicated that the
data do not require any soft blackbody component (see Table~\ref{tab:xfits}). However, as seen in Figure~\ref{fig:0729xspec}, the residuals to the bremsstrahlung in the pn
data show evidence for a broad line emission feature at 0.57~keV.  In the inset we show a fit to the line (with a linear scale).  Given the
low S/N, the significance of this feature remains questionable.  We did check whether it was an artifact of binning, by trying binning with 4 to
7 counts per bin instead, but it appeared consistently.  A line at this energy could be due to strong O\VII\ emission; we note that a number of
strong O\VI\ lines are present in the spectra of AM Her in the extreme UV \citep{mauc98}.

\subsection{\tb}
\begin{figure}[!tb]
\resizebox{.49\textwidth}{!}{\rotatebox{0}{\plotone{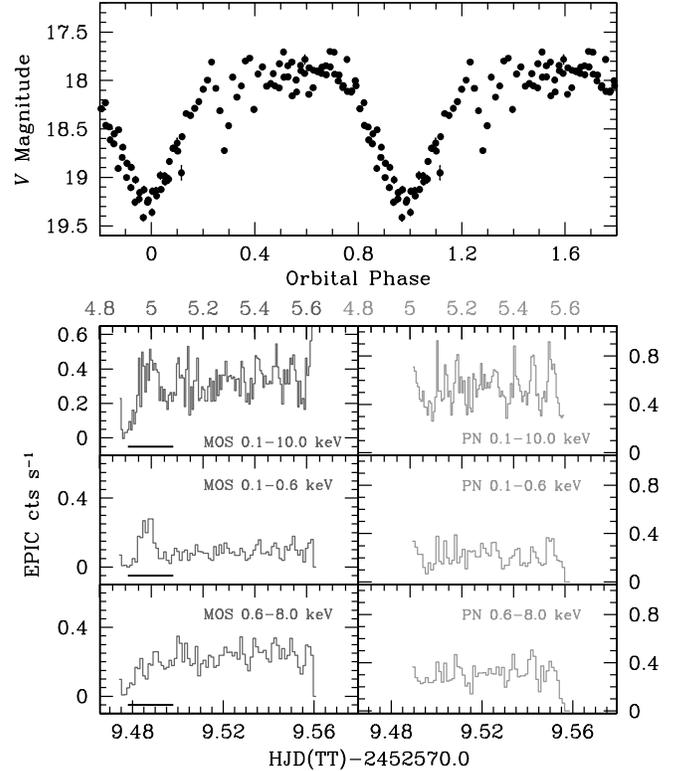}}}
\caption{Optical and X-ray lightcurves of \tb. {\em Top:} Orbital phase-folded $V$ band lightcurve. {\em Bottom:} 100s binned lightcurves from
  EPIC-MOS (left) and pn (right).  The total energy band and a soft and hard bands are shown; note that the egress from the probable eclipse is
  much more gradual at $>0.6$~keV, and the post-eclipse hump is most prominent in the soft.  Approximate phase is shown on the top axes; the
  solid bar represents the uncertainty on $T_0$ at this time.   Note HJD(UT)=HJD(TT)--64.184 s at this epoch.  \label{fig:0752lcs}}
\end{figure}

Similar to \ta, the optical light curve of \tb\ exhibits a large amplitude variation, with a prominent minimum, and
additional flux reductions \til0.3 later in phase (Fig.~\ref{fig:0752lcs}).  The PDM
analysis reveals a period of $2.74\pm0.05$ hrs.  Although the ephemeris is not well-constrained, the optical observation took place only five
cycles prior to the X-ray and we can phase the latter to within 0.1 cycles.  We therefore tentatively identify the X-ray and optical minima present at close to our
$\phi=0.0$ as due to self-eclipses.  In X-rays there appears to be a short interval of complete occultation, but unfortunately, the eclipse is right at the
start of the MOS exposure.  In the broad band optical curve, the eclipse is only partial as expected for the less localized cyclotron and
other optical emission sites.  

The optical flux dips at $\phi\sim0.3$, again considering the AM Her case, could well be due to the observer looking down along the
accretion funnel at this interval.  Moreover, immediately following the X-ray eclipse there is a distinct hump before the source settles into a
fairly steady flux level; though we note that a QPO on a 700s timescale appears towards the end of the observation.  Light curves constructed
for soft (0.1--0.6~keV) and hard (0.6--8~keV) bands, provide further insight.  The hump is much more striking
in the soft band, with the flux returning to a low (but non-zero) level for the remainder of the observation, whereas the hard band appears to come out
of eclipse and remain at a constant high level thereafter.\begin{figure*}[!tb]
\resizebox{.75\textwidth}{!}{\rotatebox{0}{\plotone{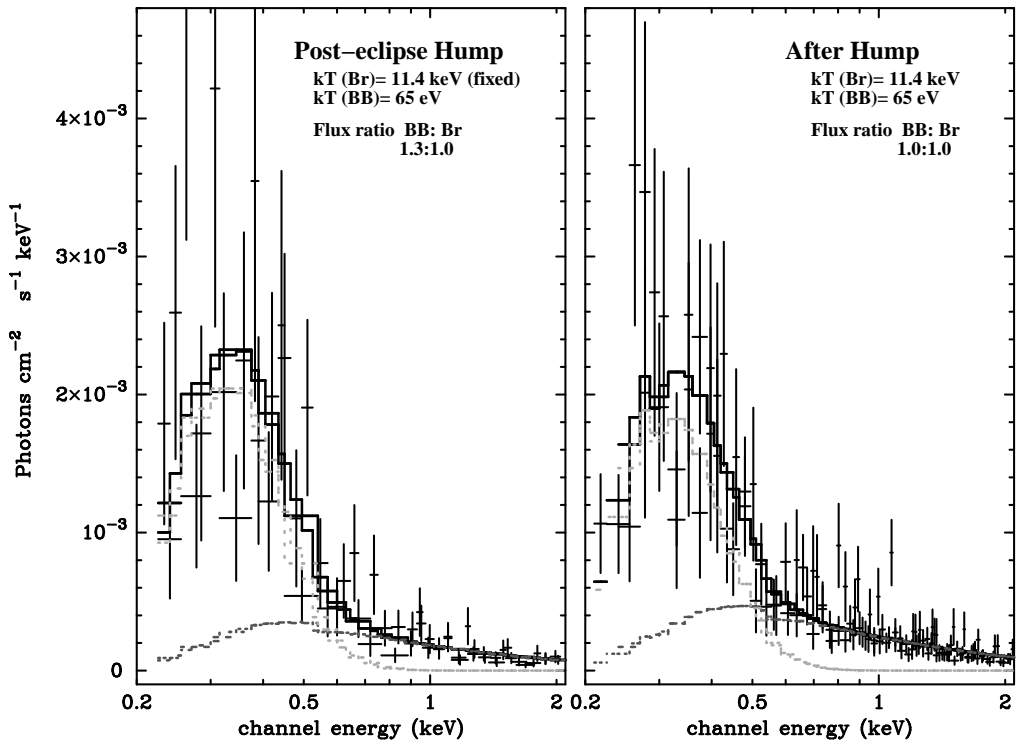}}}
\caption{Unfolded EPIC-MOS photon spectra of \tb.  {\em Left panel:} the data from the post-eclipse hump overlaid with the combined (black
  line) and separate blackbody (light grey) and
  bremsstrahlung (dark grey) contributions from the best fit model.  {\em Right panel:} the spectra, model and components for the interval following the post-eclipse hump. \label{fig:0752mosspec}}
\end{figure*}

\begin{figure}[!tb]
\begin{center}
\leavevmode 
\resizebox{.48\textwidth}{!}{\rotatebox{-90}{\plotone{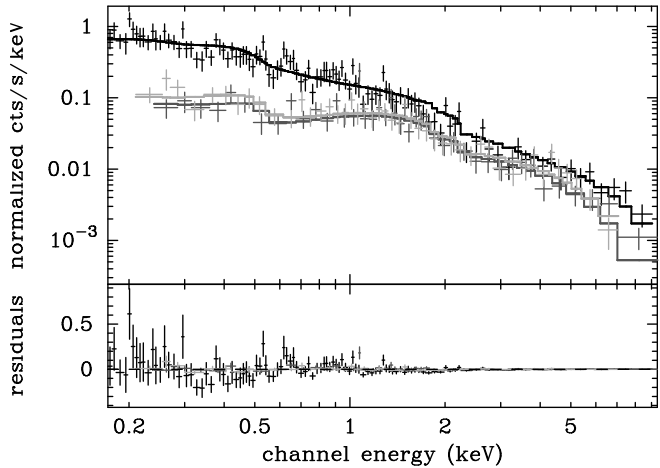}}}
\caption{EPIC spectra of \tb\ for the interval following the post-eclipse hump.  The best fit  blackbody and
  bremsstrahlung model is over-plotted and residuals are shown below.\label{fig:0752xspec}}
\end{center}
\end{figure}

We also extracted phase resolved spectra to examine the changes in greater detail (see Table~\ref{tab:xfits}).  For the interval after the hump (HJD$>2452579.495$), we fit a
basic Br + BB model (Fig.~\ref{fig:0752xspec}) and found that in this system, the BB
component is very significant; with an {\em F}-test giving probabilities $>99.99$\% for either the LR or RR (though the parameters are naturally
better constrained for the former).  Indeed, the BB contributes half of the unabsorbed flux in the 0.01--10.0~keV range, though we caution that the
exact fraction is model dependent and could be a factor of 2 different.

During the hump, we were restricted to the MOS data alone, and chose to fix $kT_{\rm Br}$ at the value found previously in the combined
MOS + pn  after-hump dataset.  The fit yielded $kT_{\rm BB}=65$ eV in both cases; hence, the change in the BB flux is not due to a change in
temperature, but rather to differing normalizations.    This suggests that as \tb\ emerges from eclipse, we can see the BB emission region fully
before it is once again obscured, most likely by the accretion column; consistent with our model for the optical lightcurve.

\subsection{\tc}

\begin{figure}[!tb]
\resizebox{.48\textwidth}{!}{\rotatebox{0}{\plotone{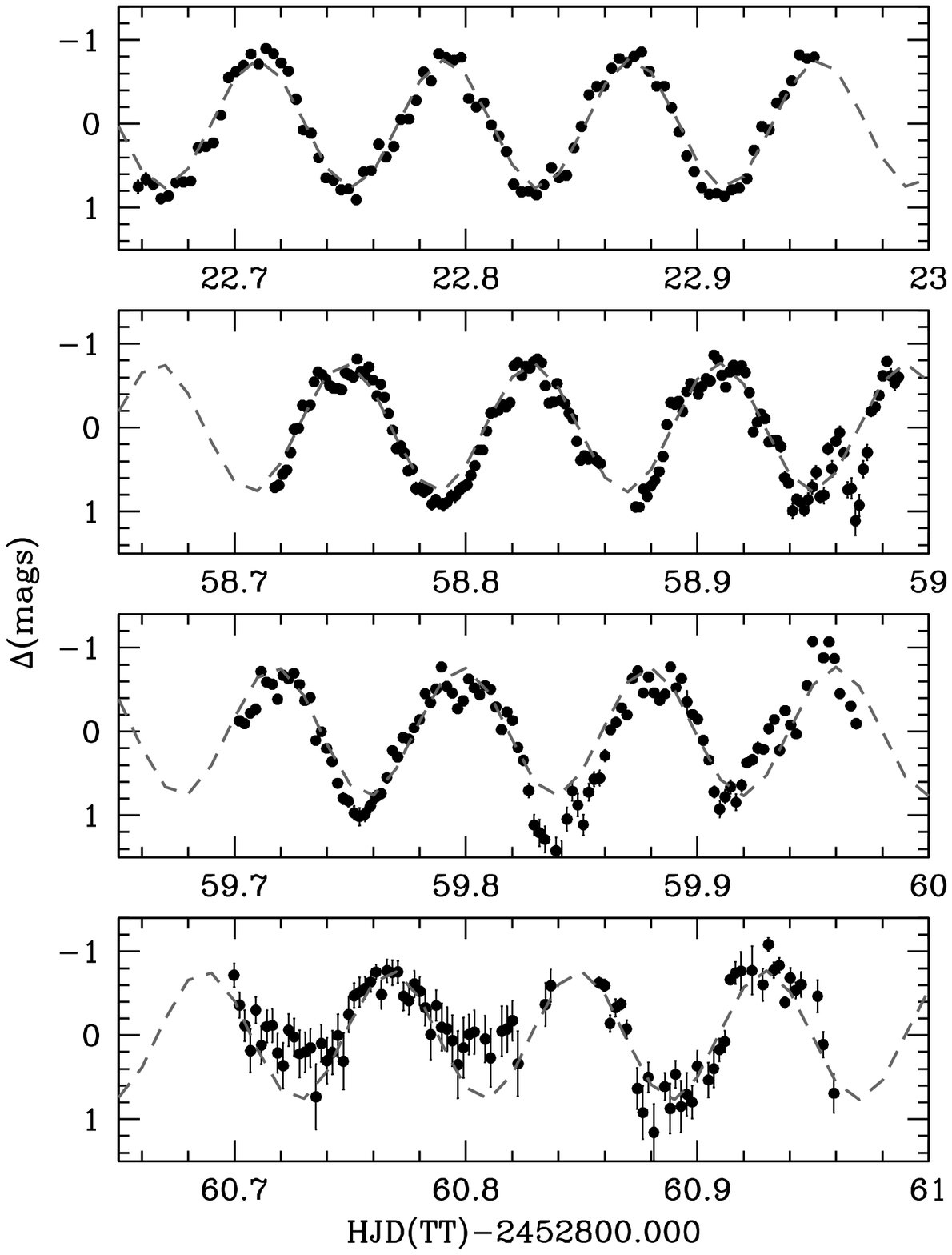}}}
\caption{Differential photometry of SDSS~J1700 from NOFS and MRO during 2003 July--August.  The data for each night have been detrended to
  emphasize the the distinct orbital modulation. The bottom panel shows a $B$-band lightcurve, the others used no filter.  Note HJD(UT)=HJD(TT)--64.184 s at this epoch. \label{fig:1700opt}}
\end{figure}

\begin{figure}[!tb]
\resizebox{.48\textwidth}{!}{\rotatebox{0}{\plotone{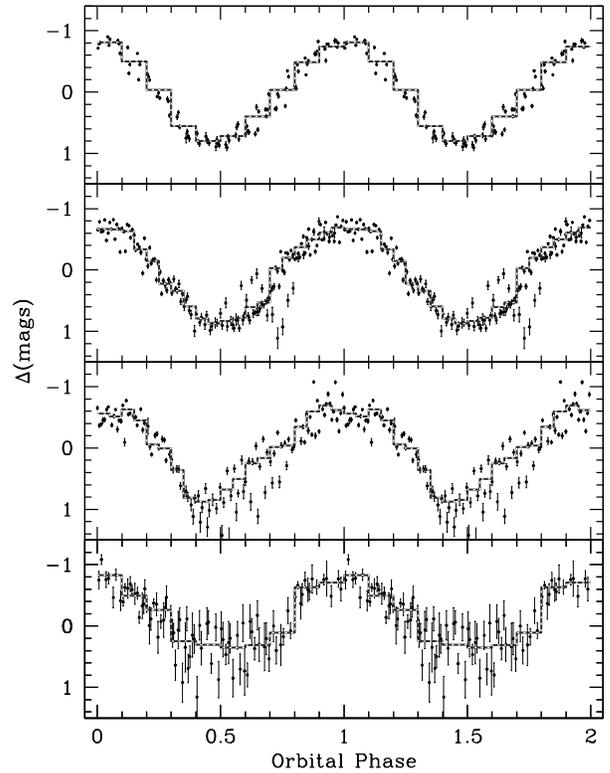}}}
\caption{The data from Figure~\ref{fig:1700opt} phase folded and binned according the best fit ephemeris.  Although close to sinusoidal the
  binned data in particular emphasizes the changing deviations. \label{fig:1700bfs}}
\end{figure}

\begin{figure}[!tb]
\resizebox{.49\textwidth}{!}{\rotatebox{0}{\plotone{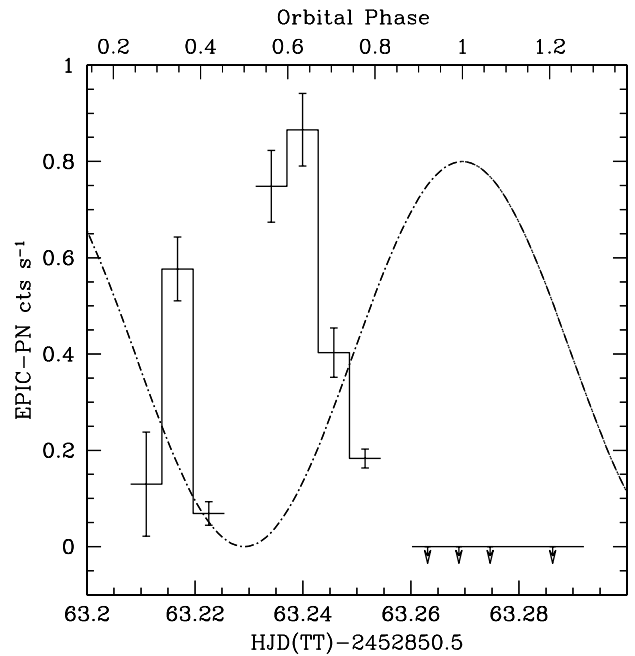}}}
\caption{EPIC-pn X-ray lightcurve of \tc.  The orbital phase is indicated on the top axis, and a sinusoid showing the phasing of the optical
  modulation is over-plotted to guide the eye to the fact that the X-ray variation is in anti-phase.  Note HJD(UT)=HJD(TT)--64.184 s at this epoch.\label{fig:1700xlc}}
\end{figure}

In Figure~\ref{fig:1700opt} we present the four nights of optical data, with the best fit sinusoid over-plotted.  The morphology of the orbital
modulation is indeed close to sinusoidal, although as illustrated in the phase folded and binned plots (Fig.~\ref{fig:1700bfs}), the exact
shape changes subtly from night to night.  Comparison to the 2001 August lightcurves presented in \citet{szko03} shows more conspicuous differences, at
that time the shape comprised broad maxima and narrow minima, in contrast to the 2003 morphology.  To derive an ephemeris, we first used a modified discrete Fourier transform, the Lomb-Scargle Periodogram
\citep{scar82}, to determine the approximate best period.  This was then refined by cycle counting ($O-C$ method), and finally by fitting a sinusoid
model to the entire dataset, yielding: \[
{\rm HJD(TT)}_{max} = 2452859.0103(3) + 0.08080175(8)\times n\]

where ${\rm HJD(TT)}_{max}$ is the time of maximum light, and the parenthetical values indicate the estimated $1\sigma$ uncertainties in the final digits.  With this precision we are easily able to phase the X-ray lightcurve (Fig.~\ref{fig:1700xlc}).

Unfortunately, during this \sat\ observation, there were technical problems, hence although we are able to accumulate data from all the small
intervals of CCD livetime, the MOS data were completely unusable for timing and only parts of the pn could be used to reliably construct a
light curve.  However, given the limited coverage it does seem that the X-ray flux is modulated in anti-phase to the optical.

There are at least two alternate explanations for the X-ray and optical lightcurves.  In the first, like \ta\ and \tb\ a single accretion pole is
visible in \tc.  The optical modulation is then a result of our viewing of the cyclotron emission, with the minimum occurring when we view down the
accretion funnel. As narrower cyclotron beaming occurs at shorter wavelengths we then obtain the broader minimum in the $B$ band
\citep[see][]{bonn85}.  In contrast, the X-ray emission reaches a maximum when the normal to the accretion spot is close to our line-of-sight.
The second model invokes two visible accretion poles, and as is the case during the ``reversed'' state of AM Her, the optical emission
is dominated by one, and the X-ray by the other.  The X-ray/optical minima are then due to total/partial self-eclipses respectively.  Only longer
contemporaneous observations of \tc\ will be able to resolve this issue.

\begin{figure}[!tb]
\resizebox{.48\textwidth}{!}{\rotatebox{-90}{\plotone{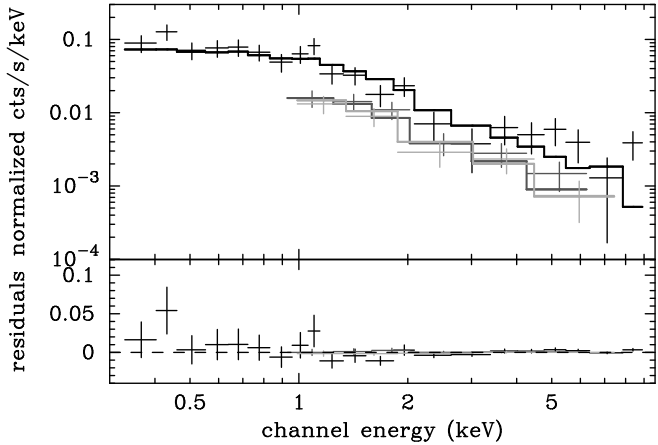}}}
\caption{EPIC spectra of \tc.  The best fit kT=7~keV MEKAL model is over-plotted and residuals are shown below.\label{fig:1700xspec}}
\end{figure}

As a consequence of the limited live time, the X-ray spectrum is of low S/N.  We fitted a variety of models to the RR and LR (see Table~\ref{tab:xfits}).  In no case was a
soft BB component (fixed at 40 eV) required.  For the Br we found a very poorly constrained temperature of $8^{+9}_{-3}$~keV.  A single
temperature MEKAL fit the data equally as well, but with a slightly better constrained $kT=7^{+5}_{-2}$~keV (see Fig.~\ref{fig:1700xspec}).  In summary, the spectrum of \tc\ is typical of a polar, with a plasma $T\sim10$~keV, but no additional soft component is required.

\section{Conclusions}
The \sat\ and optical observations presented here provide useful constraints on the accretion characteristics of \ta, \tb\ and \tc.

In terms of accretion geometries, we find \ta\ is likely a single pole accretor, exhibiting a complex optical lightcurve due to the combined
effects of cyclotron beaming and partial occultation of the emitting region. The X-ray variation, consisting of a single high flux interval
lasting perhaps only 0.2 in phase, could either be a X-ray flare event due to a sudden, short increase in the mass transfer rate from the secondary,
or the X-ray emitting region simply coming out from self-eclipse.  In any case, the X-ray spectrum at this time is most unusual, being
well-modeled by the usual hard bremsstrahlung, but also requiring a strong emission line at 0.57~keV.  This line is probably due to O\VII\
emission, and it is interesting to note that similar strong emission lines (of O\VI) are seen in the extreme UV spectra of AM Her \citep{mauc98}.

The data on \tb\ also appear to be consistent with a single pole accretor.  Both the X-ray and optical lightcurves show self-eclipses by the WD.  In the optical at
$\phi\sim0.3$ there are additional dips in the flux, possibly due to cyclotron beaming and/or obscuration effects.  In the soft X-ray, the eclipse is followed by a short rise in flux, then a long interval
at a lower level perhaps due to obscuration by the accretion column itself. This is the brightest source in X-rays, and we are able to constrain both the BB and Br components
fairly well.  We find a typical $kT_{\rm Br}=11$~keV, but a somewhat high  $kT_{\rm BB}=65$ eV, though the latter may relate to the remaining
low energy calibration uncertainties of the EPIC instruments.  Bearing this, and the uncertainty in the absorbing column which we could not
constrain, in mind we can derive a value for the soft/hard X-ray energy balance. Taking into account the geometrical effects, the X-ray scattering albedo of the
WD photosphere, $a_X=0.3$ \citep{will87}, but not the
cyclotron term (typically negligible compared to the Br) we find $L_{\rm BB}/ L_{\rm Br}\approx(\pi f_{\rm BB}(1-a_x) d^2)/(2\pi f_{\rm
  Br}(1+a_x) d^2) = 0.35-0.5$.  This is well-within the range found in a recent survey of polars by \citet{rams04a}, and in agreement with the
basic picture of X-ray emission from radial accretion.

For \tc\ the lightcurves allow two alternate models, either a single or two pole accretor, and both are able to explain the anti-phasing of the
X-ray and optical modulations.  In the former, the cyclotron beaming minimum occurs at roughly the same phase that the accretion region appears
brightest in X-rays.  In the latter, we postulate that the optical and X-ray emission originate at different poles, and are principally
modulated by the occurrence of self-eclipses.

\acknowledgments
This work was funded by NASA \sat\ grant NAG5-12938 and NSF grant AST-02-05875 to the University of Washington.  It is based on observations
obtained with \sat, an ESA science mission with instruments and contributions directly funded by ESA member states and the USA (NASA).
   
Funding for the creation and distribution of the SDSS Archive has
been provided by the Alfred P. Sloan Foundation, the Participating
Institutions, the National Aeronautics and Space Administration, the
National Science Foundation, the U.S. Department of Energy, the
Japanese Monbukagakusho, and the Max Planck Society. The SDSS Web site
is http://www.sdss.org/.

    The SDSS is managed by the Astrophysical Research Consortium (ARC)
for the Participating Institutions. The Participating Institutions are
The University of Chicago, Fermilab, the Institute for Advanced Study,
the Japan Participation Group, The Johns Hopkins University, the Korean
Scientist Group, Los Alamos National Laboratory, the
Max-Planck-Institute for Astronomy (MPIA), the Max-Planck-Institute
for Astrophysics (MPA), New Mexico State University, University of
Pittsburgh, Princeton University, the United States Naval Observatory,
and the University of Washington.


\begin{deluxetable*}{llllllll}[!tb]
\tablewidth{0pt}
\tablecaption{Summary of the X-ray Spectral Fits\label{tab:xfits}
}
\tablehead{\colhead{SDSS J}& \colhead{Energy\tablenotemark{a}} & \colhead{Model} &\colhead{Goodness\tablenotemark{b}} & \colhead{$N_H$} & \colhead{$kT$ or $E$} & \multicolumn{2}{c}{Flux\tablenotemark{c}
	}\\
 & \colhead{range}& & \colhead{of Fit} & \colhead{$\times10^{20}$cm$^{-2}$}& & }
\startdata
0729 & RR& Bremss &54\% & 5.96 (f)\tablenotemark{d} & $10^{+13}_{-4}$~keV & 4.2  \\ 
&& BB &47\% &&40 eV (f) & 16  \\
&& + Bremss & & &30~keV (f) & \\
&& Bremss & 8\% & & 30~keV (f) & 3.8 \\
&& + Gaussian line & & &$0.56\pm0.02$~keV & 0.2 &\\
0752 &LR & BB & 36\%\tablenotemark{e} & 5.11 (f)  & $66\pm3$ eV & 18 \\
(after-hump) && + Bremss &  & & $11.4\pm1.9$~keV & 26 \\
(hump) && BB & 74\% & & $65\pm4$ eV & 18\\
       && +Bremss & && 11.4 (f) & 18\\
1700 & RR& Bremss & 1.0 & 2.15 & $8^{+9}_{-3}$keV &  8\\
&& BB & 0.97 & & 40 eV (f)& 14  \\
&&+Bremss & & & $12^{+14}_{-5}$~keV &  \\
&& mekal & 0.97  & & $7^{+5}_{-2}$~keV &  8\\
\enddata
\tablenotetext{a}{.LR= Large Range.  Using 0.15--10~keV for pn and 0.2--10~keV for MOS to make most of low-energy response.  RR= Restricted Range.  More
  conservative lower energy limits imposed: 0.3--10~keV for pn and 0.5--10~keV for MOS.
}
\tablenotetext{b}{For \ta\ and \tb\ where the fitting utilized Cash statistics, the goodness of fit is found via a Monte Carlo method.
  Sampling each parameter randomly within its allowed distribution $\simgt1000$ spectra were simulated, and a fit performed.  The percentage
  refers to the incidence of fits with lower C-statistic values than that of the fit to the data.  A good fit
  should have a value around 50\%.  For \tc\ the reduced $\chi^2$ is quoted.}

\tablenotetext{c}{Unabsorbed flux in the 0.01--10~keV range, in units of $10^{-13}$\ergsqcmsec, including the correction for the 70\% encircled energy fraction.}
\tablenotetext{d}{(f) indicates that the parameter was frozen at this value.}

\tablenotetext{e}{For a single-component Bremsstrahlung model the percentage was 100\%, clearly indicating the need for the blackbody.}
\end{deluxetable*}

\end{document}